\documentclass[runningheads]{llncs}
\usepackage{graphicx}
\usepackage{cleveref}
\usepackage{booktabs}
\usepackage{listings}
\begin{document}
\title{Detect Llama - Finding Vulnerabilities in Smart Contracts using Large Language Models}
\titlerunning{Detect Llama}
%
\author{Peter Ince\inst{1} \and
Xiapu Luo\inst{2} \and
Jiangshan Yu\inst{3} \and
Joseph K. Liu\inst{1}\and
Xiaoning Du\inst{1}}
\authorrunning{Ince et al.}
%
\institute{Monash University, Clayton, Australia
\email{\{peter.ince1,jospeh.liu,xiaoning.du\}@monash.edu}\\
 \and
The Hong Kong Polytechnic University, Hung Hom, Hong Kong \\
\email{csxluo@comp.polyu.edu.hk} \\
\and
University of Sydney, Darlington, Australia \\
\email{Jiangshan.yu@sydney.edu.au}}
\maketitle              
\begin{abstract}
In this paper, we test the hypothesis that although OpenAI's GPT-4 performs well generally, we can fine-tune open-source models to outperform GPT-4 in smart contract vulnerability detection.

  We fine-tune two models from Meta's Code Llama and a dataset of 17k prompts, Detect Llama - Foundation and Detect Llama - Instruct, and we also fine-tune OpenAI's GPT-3.5 Turbo model (GPT-3.5FT).

  We then evaluate these models, plus a random baseline, on a testset we develop against GPT-4, and GPT-4 Turbo's, detection of eight vulnerabilities from the dataset and the two top identified vulnerabilities - and their weighted F1 scores.

  We find that for binary classification (i.e., is this smart contract vulnerable?), our two best-performing models, GPT-3.5FT and Detect Llama - Foundation, achieve F1 scores of  $0.776$ and $0.68$, outperforming both GPT-4 and GPT-4 Turbo, $0.66$ and $0.675$.

  For the evaluation against individual vulnerability identification, our top two models, GPT-3.5FT and Detect Llama - Foundation, both significantly outperformed GPT-4 and GPT-4 Turbo in both weighted F1 for all vulnerabilities ($0.61$ and $0.56$ respectively against GPT-4's $0.218$ and GPT-4 Turbo's $0.243$) and weighted F1 for the top two identified vulnerabilities ($0.719$ for GPT-3.5FT, $0.674$ for Detect Llama - Foundation against GPT-4's $0.363$ and GPT-4 Turbo's $0.429$).
\end{abstract}

\section{Introduction}
Over the past few years, we have seen Decentralised Finance (DeFi) expand over chains and grow its usage - often measured in Total Value Locked (TvL). At its peak in 2022, DeFi's TvL over all chains reached almost USD\$250B, resting at approximately USD\$54B as at November 2023\cite{defillama_defillama_2023}.
With this influx of money comes attention to the protocols and networks of bad actors and hackers. Over the past few years, blockchain networks have seen 148 exploits worth approximately USD\$4.28B\cite{chainsec_comprehensive_2023}.

These continued attacks highlight the need for more tools to detect vulnerabilities in smart contracts quickly with as few false positives as possible.
There are many great tools for automated smart contract vulnerability detection; however, each category has its own challenges;
\begin{itemize}
    \item \textbf{Static Analysis tools} are fast but often produce false positives
    \item \textbf{Dynamic analysis tools} including fuzzing and static analysis tools tend to be more accurate but can take a significant amount of time to identify vulnerability
\end{itemize}

Consequently, there is much need for a tool that encompasses the best of both static analysis and fuzzing/symbolic execution tools - that is both fast and reduces the capturing of false positives in results.

We have seen Large Language Models such as OpenAI's GPT-4\cite{openai_gpt-4_2023} perform relatively well for few-shot learning when it comes to detection classification of the vulnerable state of Solidity smart contracts\cite{david_you_2023}.

We hypothesize that by training the most performant open-source code-based Large Language Model available with labelled Solidity smart contract vulnerabilities, we can outperform GPT-4 and offer a middle tool in-between static analysis and fuzzing/symbolic execution.

In this paper, we leverage Meta's Code Llama models\cite{roziere_code_2023} and fine-tune them on a dataset of 17,000 prompts created from a dataset of 9,252 labelled smart contracts\cite{yashavant_scrawld_2022} and produce two open-source models based on the Code Llama 34b parameter Foundation and Instruct tuned models.

We also fine-tune OpenAI's GPT-3.5 Turbo model\cite{peng_gpt-35_2023} with a subset of 4,000 prompts, and create a random baseline for comparison.

We then create a custom test set, compare the three fine-tuned models with the random baseline, GPT-4 and GPT-4 Turbo and analyse the results.

We find that while, in general, all the fine-tuned models outperform GPT-4 and GPT-4 Turbo - the fine-tuned GPT-3.5 Turbo outperforms all of the other models with a weighted F1 Score of $0.61$ on all eight vulnerabilities and $0.71$ on the two most accurate vulnerabilities, with the Code Llama 34b Foundation based model performing slightly less well with a weighted F1 score of $0.586$ and $0.674$ respectively.

\subsection{Our contributions}
In this paper, we make the following contributions to the field of smart contract security;
\begin{itemize}
    \item We release some of the first open-source Large Language Models for specialisation as a smart contract vulnerability detection tool\cite{ince_detect_2023,ince_detect_2023-1} (a fine-tuned version of Code Llama 34b models)
    \item We fine-tune and evaluate GPT-3.5 Turbo as a smart contract vulnerability detection tool
    \item We evaluate GPT-4 and GPT-4 Turbo as smart contract vulnerability detection tools and show that both open-source models and GPT-3.5 Turbo can be fine-tuned to significantly outperform GPT-4 and GPT-4 Turbo on specific detection tasks
    \item We release both our open-source models and the prompt sets used for both training\cite{ince_smart_2023} and evaluation\cite{ince_peterdouglasdetect-llama-evaluation_2024} to allow for future research to build upon our work
\end{itemize}

\subsection{Structure}
The remainder of this paper comprises the following sections; in \cref{sec:background}, we provide some necessary background to give context and understanding to our research. Section \ref{sec:related} covers related research to our work, and \cref{sec:approach} details our approach for preparing our dataset and training prompts, and fine-tuning our models.

In \cref{sec:eval}, we share the process and details of our model evaluation, and in \cref{sec:discuss}, we discuss some of the improvements that could be made to our models and dataset and future work.

\section{Background}\label{sec:background}
In this section, we provide necessary background information on \textit{Detecting Vulnerabilities in Smart Contracts}, \textit{Detecting Vulnerabilities in Smart Contracts with AI}, and \textit{Generate Pre-trained Transformers (GPTs)}.

\subsection{Detecting Vulnerabilities in Smart Contracts}

For almost as long as smart contracts have existed on Ethereum, there have been people attempting to exploit vulnerabilities in the code for financial gains, such as the DAO Attack in 2016 that caused a hard fork of both the Ethereum chain and community\cite{siegel_understanding_2016}.

As a result, we have seen both the rise of smart contract auditors (typically firms or individuals that specialise in testing and analysing for specific vulnerabilities in smart contracts) alongside automated tools that assist with identifying vulnerabilities.

There are two primary categories of automated tools; these tools are often used with manual smart contract testing to assist in the security assurance of smart contracts.

\subsubsection{Static analysis} Static analysis tools, such as Slither\cite{feist_slither_2019} and SmartCheck\cite{tikhomirov_smartcheck_2018}, work via analysing the code for exploits without executing the smart contract\cite{feist_slither_2019}. Some static analysis tools, such as \cite{feist_slither_2019} and \cite{tikhomirov_smartcheck_2018}, may also translate the source code into some form of intermediate representation to simplify the representation and analysis while still maintaining overall semantics.

\subsubsection{Dynamic analysis} Dynamic analysis tools include both symbolic execution tools, such as Oyente\cite{luu_making_2016}, Osiris\cite{torres_osiris_2018} and Mythril\cite{consensys_mythril_2023}; and fuzzing tools, such as ItyFuzz\cite{shou_ityfuzz_2023} and ConFuzzius\cite{torres_confuzzius_2021}.

Dynamic analysis tools work by executing the contract in various ways; symbolic execution converts the smart contract code to representative symbols and uses a constraint solver, such as Z3\cite{de_moura_z3_2008}, to determine whether or not there are any vulnerabilities.
Fuzzzers, or fuzz-testing tools, use various forms of trace analysis, taint analysis, input mutation, or other techniques to generate input transactions to the deployed smart contract.

\subsection{Detecting Vulnerabilities in Smart Contracts with AI}
There have been several tools that have used various forms of Artificial Intelligence, or Machine Learning, to identify vulnerabilities in smart contracts.
Most AI/ML tools for detecting vulnerabilities in smart contracts start with a dataset of labelled smart contracts. These smart contracts are identified and labelled using static and dynamic analysis tools, manual identification, or a combination.

Some examples of labelled datasets are \cite{liu_rethinking_2023} and \cite{yashavant_scrawld_2022}.
In \cite{lutz_escort_2021}, Lutz et al used a Deep Neural Network to identify six vulnerability types with an F1 score of 96\% and were able to use transfer learning to identify new vulnerabilities with an average F1 score above 90\%\cite{lutz_escort_2021}.

Also, in \cite{tann_towards_2019}, Tann et al use Long Short Term Memory (LSTM) and train their model on their own dataset, achieving fast, large-scale analysis of contracts with a 99\% accuracy rate\cite{tann_towards_2019}.

\subsection{Generative Pre-trained Transformers (GPTs)}
Generative Pre-trained Transformers (GPTs) are language models that are pre-trained (i.e., unsupervised) on a large corpus of information, often crawled from the internet for text and general understanding or taken from open-source code repository sites. Open-source Language Models, such as the StarCoder models\cite{li_starcoder_2023}, also choose to open-source the repositories used in the pre-training phase\cite{Kocetkov2022TheStack}.

GPT language models can then be "fine-tuned" on specific prompt and response sets. The process of fine-tuning alters some of the model parameters to fit the data provided in the prompts; an example of this is OpenAI's GPT-3.5 (a version of which we use for our evaluation), which is a version of GPT3\cite{brown_language_2020} that has been fine-tuned using RLHF (Reinforcement Learning from Human Feedback)\cite{ouyang_training_2022}.

Innovations such as LORA (Low-Rank Adaption)\cite{hu_lora_2021} reduce the overhead of memory required for the fine-tuning of large language models by selectively modifying a small percentage of training parameters instead of modifying them all\cite{hu_lora_2021}. QLoRA (Quantized Low-Rank Adaption) builds on the world by Hu et al in \cite{hu_lora_2021} by using 4-bit quantization to further reduce the memory overhead\cite{dettmers_qlora_2023}.

The current state-of-the-art GPT model is OpenAI's GPT-4\cite{openai_gpt-4_2023} and GPT-4 Turbo Preview\cite{openai_new_2023}(sometimes referred to simply as GPT-4 Turbo in this paper), which were not available for fine-tuning at the time of writing this paper and is closed-source, so can only be accessed through OpenAI's API.

\section{Related work}\label{sec:related}
\subsection{LLMs for vulnerability detection}
Since the release of ChatGPT by OpenAI in late 2022, we have seen an invigoration of interest in using Large Language Models for various use cases.

For smart contract vulnerability detection with LLMs, we have seen two approaches; the first, by David et al, identified a set of historically vulnerable smart contracts and applied state-of-the-art (SOTA) LLMs as few-shot learners, namely GPT-4-32k from OpenAI and Claude from Anthropic, to detect vulnerabilities in those historical smart contracts. David et al also created some smaller smart contracts for testing with specific vulnerabilities inserted\cite{david_you_2023}.
In \cite{david_you_2023}, David et al found that while the best-performing model, GPT-4-32k, was able to detect vulnerable smart contracts with a True Positive rate of $78.7\%$, however, the rate of correct vulnerability identification was only $40\%$\cite{david_you_2023}.

The other approach by Gai et al trained a Large Language Model on a dataset of over 68 million transactions focusing on previously compromised DeFi smart contracts\cite{gai_blockchain_2023}. Gai et al's trained Language Model becomes a part of their intrusion detection system, \textit{BlockGPT}\cite{gai_blockchain_2023}, which seeks to identify abnormal transactions while they are in the mempool (i.e., before the application processes them), so that the protocol can be paused before an attack can be executed on the smart contract\cite{gai_blockchain_2023}.

Another work that uses GPT for smart contract vulnerability detection is \cite{hu_large_2023} by Hu et al. Hu et al propose GPTLens, a system for vulnerability detection using open-ended prompting with the addition of a two-step $auditor \rightarrow critic$ process that analyses the detected vulnerabilities and ranks them based on their $correctness$, $severity$ and $profitability$ rating\cite{hu_large_2023}.

To further compare our smart contract vulnerability detection models, we use a modified version of GPTLens\cite{hu_large_2023} and a version of their critic analysis technique to validate identified vulnerabilities.
 
\section{Approach}\label{sec:approach}

\subsection{Dataset selection and processing}
For our dataset, we wanted it to meet two criteria;
\begin{enumerate}
    \item It should have a large number of smart contracts with vulnerability labels for training
    \item It should allow the process to be able to generate our test set to validate our models
\end{enumerate}

The dataset that had the largest amount of Ethereum Solidity smart contracts with vulnerability labels that we found in our investigation was \textit{ScrawlD: A Dataset of Real World Ethereum Smart Contracts Labelled with Vulnerabilities}\cite{yashavant_scrawld_2022} by Yashavant et al.

In \cite{yashavant_scrawld_2022}, Yashavant et al use a suite of 5 different tools to identify vulnerable smart contracts using a majority vote approach across 8 different vulnerabilities\cite{yashavant_scrawld_2022}.

The latest update to the dataset from \cite{yashavant_scrawld_2022} by Yashavant et al includes  9,252 smart contracts, 5,364 of which contain at least one vulnerability\cite{yashavant_scrawld_2023}.

\subsubsection{Processing}
The dataset from \cite{yashavant_scrawld_2022},\cite{yashavant_scrawld_2023} contains only the Ethereum addresses of the smart contracts. 

Therefore, our process to prepare the smart contracts is as follows;
\begin{enumerate}
    \item Download the smart contract code from EtherScan's Verified Smart Contract API\cite{etherscanio_etherscanio_nodate}
    \item Remove comments and additional new lines from the smart contracts
    \item Add vulnerability label data to each smart contract record
\end{enumerate}

For training our models, our context window is limited (as discussed in \cref{sec:models}); therefore, we measure the number of tokens using the GPT2 Tokenizer, sort the records and exclude the top 750 smart contracts (those over 7340 tokens in length).

Our analysis showed that most token lengths appear to be in the range $(0,\leq7340)$, with some outliers far beyond the range (with the highest having more than 100,000 tokens).

\subsection{Prompt strategy}\label{sec:prompt_strategy}
Once we have the processed records for the smart contract, including their vulnerabilities, we are able to put them together with the prompts.

We are using the \textit{Alpaca Instruct}\cite{taori_alpaca_nodate} prompt style for our prompting of open source models.

We took each of the smart contracts and their corresponding vulnerabilities and turned them into prompts in two styles;
\begin{itemize}
    \item The first style prompt is focused on the generation of smart contracts; both those with at least 1 vulnerability (see listing \ref{lst:detection_has_vuln}) and those without any detected vulnerability (see listing \ref{lst:detection_no_vuln}).
    \item The second prompt style is focused on the detection of vulnerable smart contracts; as with the previous style, with at least 1 vulnerability (see listing \ref{lst:generate_has_vuln}) and no detected vulnerabilities (see listing \ref{lst:generate_no_vuln}).
\end{itemize}

Our goal was to add an additional layer of data to the model around what constitutes the particular types of vulnerability to the models by approaching the same vulnerability from two perspectives in the training prompts.

\begin{lstlisting}[caption={Instruction-input style prompt for detection when an example contains 1 or more vulnerabilities},label={lst:detection_has_vuln}]
{"instruction": "You are an expert AI system trained to assist with smart contract security by analysing Solidity smart contracts for vulnerabilities.", 
"input": "Please analyse the following smart contract for vulnerabilities: <smart contract code>", 
"output": "The provided contract has 1 or more of the following vulnerabilities:
<Listed Vulnerabilities in the format SWC-ID - Vulnerability name>"}
\end{lstlisting}

\begin{lstlisting}[caption={Instruction-input style prompt for detection when an example contains no vulnerabilities - only output difference shown},label=lst:detection_no_vuln]
{"instruction": "...", 
"input": "...", 
"output": "The provided smart contract has none of the following vulnerabilities:
        <All Eight included vulnerabilities in the format SWC-ID - Vulnerability name>"}
\end{lstlisting}

\begin{lstlisting}[caption={Instruction-input style prompt for generation when a smart contract contains 1 or more vulnerabilities},label=lst:generate_has_vuln]
{"instruction": "You are an expert AI system trained to assist with smart contract security by generating vulnerable and non-vulnerable smart contracts on the Ethereum blockchain, written in Solidity.", 
"input": "Generate an Ethereum smart contract written in Solidity that has 1 or more of the following vulnerabilities:
<All Eight included vulnerabilities in the format SWC-ID - Vulnerability name>", 
"output": "<smart contract code>"}
\end{lstlisting}

\begin{lstlisting}[caption={Instruction-input style prompt for generating a smart contract when example contains no vulnerabilities - only input difference shown. Vulnerabilities listed from \cite{yashavant_scrawld_2022}},label=lst:generate_no_vuln]
{"instruction": "...", 
"input": "Generate an Ethereum smart contract written in Solidity that has none of the following vulnerabilities:
        <All Eight included vulnerabilities in the format SWC-ID - Vulnerability name>", 
"output": "<smart contract code>"}
\end{lstlisting}

\subsection{Model selection and training}\label{sec:models}
One of the challenges inherent in training a Large Language Model for detecting and generating smart contract vulnerabilities is the context window (the number of tokens allowed in the input) and the total number of tokens (both input and output). These challenges exist because smart contracts vary wildly in length. Therefore, a language model must have a relatively large context window to be useful for vulnerability detection.

However, most state-of-the-art open-source Large Language Models have had a smaller context window (usually around 2,000 tokens, as with the initial version of WizardCoder\cite{luo_wizardcoder_2023} by Luo et al), especially those constrained by the cost of the hardware (or cloud resource rental) associated with training LLMs.

Some open-source models have a larger context window, such as the StarCoder series of models with a context window of 8,000 tokens\cite{li_starcoder_2023}. However, the model did not perform as well on evaluation metrics as other open-source models\cite{li_starcoder_2023}.

Open-source LLMs made an evolutionary leap when Meta released their collection of 
\textit{Code Llama} models\cite{roziere_code_2023}. With \cite{roziere_code_2023}, Rozier et al released a series of models - a foundation (aka a base), a Python-tuned, and an Instruct-tuned model. These models were released in three sizes: 7 billion, 13 billion, and 34 billion parameters\cite{roziere_code_2023}. Not only did these models outperform many other LLMs on benchmarks like HumanEval (such as Luo et al's StarCoder models\cite{li_starcoder_2023}, but they were also trained on a larger input context window of 16,000 tokens and supported up to 100,000 token context windows\cite{roziere_code_2023}. 

This extended content window and improved performance made Code Llama the right base model for us.
\subsubsection{Code Llama}
For fine-tuning of the Code Llama models, we used the dataset created as described in \cref{sec:prompt_strategy}. The final training dataset was 17,000 records in length.

For training, we used a context window of 7500 tokens, three epochs, ten warm-up steps and 20 eval steps; and to allow us to train a larger model on less GPU hardware, we used QLORA\cite{dettmers_qlora_2023} and Flash Attention V2\cite{dao_flashattention-2_2023}. 

\subsubsection{GPT-3.5 Finetune}
For fine-tuning of GPT-3.5 Turbo\cite{peng_gpt-35_2023}, we used a smaller dataset of 4,000 records (primarily a cost constraint). The training featured only the prompts for detection featured in \cref{sec:prompt_strategy} and used the ChatGPT prompt style\cite{peng_gpt-35_2023} instead of the Alpaca Instruct\cite{taori_alpaca_nodate} style.

In total, we trained a total of 16,906,389 tokens over three epochs.

\section{Evaluation}\label{sec:eval}
To evaluate the effectiveness of the models, we must create our own test set.

\subsection{Building the test data}
As some of the tools used by Yashavant et al in \cite{yashavant_scrawld_2022} had not been updated for later versions of the Solidity compiler, all of the smart contracts in the test set had to be $0.4.x$ (i.e. - the version of Solidity used must be $\geq 0.4.0$ and $\leq 0.4.26$).

Given this requirement, we analysed the data on Ethereum to find the top open-source smart contracts using version $0.4.x$ and downloaded approximately 600 smart contracts.

We then individually ran all of the tools used by \cite{yashavant_scrawld_2022} on the smart contracts;
\begin{enumerate}
    \item Osiris\cite{torres_osiris_2018}
    \item Oyente\cite{luu_making_2016}
    \item Mythril\cite{consensys_mythril_2023}
    \item Slither\cite{feist_slither_2019}
    \item SmartCheck\cite{tikhomirov_smartcheck_2018}
\end{enumerate}

Some modifications had to be made to account for different solidity versions in the $\geq 0.4.0$ and $\leq 0.4.26$ range.

We then processed the smart contract files using the files and processes by Yashavant et al in \cite{yashavant_scrawld_2022} and \cite{yashavant_scrawld_2023}.

\subsection{Setting a random baseline}

We then created a random baseline. Each smart contract was randomly assigned between 0 and 4 of the smart contract vulnerabilities from \cite{yashavant_scrawld_2022}.

\subsection{Implementation}

Gathering the results of the tests involved us searching for vulnerabilities with each of the models we are testing.

For the \textit{Detect Llama} models based on Meta's Code Llama models\cite{roziere_code_2023} we use the input style shown in listing \ref{lst:detection_has_vuln} with the Alpaca Instruct\cite{taori_alpaca_nodate} prompt style.

For our fine-tuned GPT-3.5 Turbo, we use the same input style as shown in listing \ref{lst:detection_has_vuln} with the ChatGPT prompt style\cite{peng_gpt-35_2023}.

To perform the Zero-shot GPT-4 and GPT-4 Turbo analysis, we use the prompt shown in listing \ref{lst:gpt4_prompt} - the prompt uses learnings from \cite{kojima_large_2022} (part of the prompt is \textit{Think step by step}\cite{kojima_large_2022} - in conjunction with the using the function calling feature\cite{eleti_function_2023} to structure the analysis responses efficiently as JSON.

\begin{lstlisting}[caption={Prompt used for GPT-4 Zero-shot analysis -  with prompt tuning seen in \cite{kojima_large_2022}},label=lst:gpt4_prompt]
You are a world renown smart contract auditor. You must analyze Ethereum smart contracts to detect exploits and develop example code to test the exploit to validate it. You are able to utilize fuzzing techniques to locate and fix weaknesses in the contracts, while also understanding the concepts of cryptography, blockchain technology, and secure coding practices. 
The specific exploits you MUST search for in each smart contract are;
<All Eight included vulnerabilities in the format SWC-ID - Vulnerability name>

Rules you MUST follow:
- Be brief and to the point
- Think step by step
- Try your best to avoid false positives in exploit identification
- Provide the code vulnerable code from the smart contract with line numbers
- "Status" should be only "No Exploit" or "Exploit Found"
\end{lstlisting}

\subsection{Alternate technique evaluation}
\subsubsection{GPTLens}
To assist in evaluating our models, we also compare them against results generated using techniques from GPTLens, developed by Hu et al in \cite{hu_large_2023}.

However, as the auditing prompt in GPTLens is designed to be open-ended while searching for vulnerabilities\cite{hu_large_2023}, we must add some specifications around the vulnerabilities we are searching for.

In \cite{hu_large_2023}, Hu et al find the best results with one auditor and one critic, finding up to 3 vulnerabilities. 

Each smart contract is processed as follows;
\begin{itemize}
    \item Smart contract uses the auditor prompt from \cite{hu_large_2023}, modified to search within the 8 vulnerabilities defined in the dataset\cite{yashavant_scrawld_2022}, returning the top 3 vulnerabilities.
    \item The few-shot critic prompt is run against the audit response and graded on a scale of 0-10 for \textit{correctness}, \textit{severity} and \textit{profitability}\cite{hu_large_2023}.
    \item The ranking algorithm is then run to calculate a \textit{final score} based on the \textit{correctness}, \textit{severity} and \textit{profitability} ratings returned by the critic\cite{hu_large_2023}.
\end{itemize}

\subsubsection{Critic analysis}
In addition to the GPTLens style analysis, we also use the two-step process of $analysis \rightarrow critic$ proposed in \cite{hu_large_2023} to augment our Zero-shot analysis using GPT-4 and GPT-4 Turbo.

For each evaluation response from our GPT-4 and GPT-4 Turbo vulnerability detection, we use our critic prompt set; the system prompt is shown in listing \ref{lst:gpt4_critic} with the individual prompt shown in listing \ref{lst:gpt4_critic_vuln}. Note that our critic prompt also uses the \textit{Think step by step} from \cite{kojima_large_2022}.

\begin{lstlisting}[caption={System prompt used for GPT-4/GPT-4 Turbo Critic Analysis with prompt tuning from \cite{kojima_large_2022}},label=lst:gpt4_critic]
The vulnerabilities and listed code combinations are likely to contain mistakes. As a harsh vulnerability critic, your duty is to scrutinize the exploit listed and associated code and evaluate the correctness and severity of given vulnerabilities and associated reasoning and provide a 'confirm' or 'reject' response with detailed feedback.

Rules you MUST follow:
- Be brief and to the point
- Think step by step
- "Status" should only be 'No changes recommended' when you have not rejected any exploits identified and have not put any rejected exploits in exploits_rejected, or 'Changes recommended' if you have rejected any exploits and stored them in exploits_rejected
- "Exploits" should contain the confirmed exploits with your feedback
- "Exploits_rejected" should contain the rejected exploits with the reason for rejection
\end{lstlisting}

\begin{lstlisting}[caption={Example prompt for criticism of detected vulnerability analysis},label=lst:gpt4_critic_vuln]
please critique these exploit and code combinations for Ethereum smart contracts written in Solidity: 

======== EXPLOIT 1 ========

exploit : SWC-107 - Reentrancy 

code : Lines 138-144: 
function transfer(address _to, uint _value) public whenNotPaused {
    require(!isBlackListed[msg.sender]);
    if (deprecated) {
        return UpgradedStandardToken(upgradedAddress).transferByLegacy(msg.sender, _to, _value);
    } else {
        return super.transfer(_to, _value);
    }
} 

<continued for each exploit>
\end{lstlisting}
We evaluated the entire test set using a modified GPTLens\cite{hu_large_2023} technique. In \cite{hu_large_2023}, Hu et al calculate the \textit{final score} in addition to the \textit{correctness}, \textit{severity} and \textit{profitability} of the vulnerability.

As we only seek to determine whether the vulnerability analysis is correct (i.e., is the smart contract vulnerable), we focus our testing primarily on \textit{correctness}.

In \cref{tab:gptlens_eval}, we show our evaluation of the results from GPTLens\cite{hu_large_2023} with different parameters for inclusion of results. The results for DOS F1 and Tx-Origin FT have been excluded as they were all zero.

For the GPTLens results shown in \cref{tab:gptlens_eval}, 75 vulnerability descriptions were returned and reclassified into the eight distinct vulnerabilities, with 23 unrelated vulnerability types excluded from reporting. 

\begin{table}[h]
\centering
\resizebox{\textwidth}{!}{%
\begin{tabular}{|p{2.5cm}|p{1.5cm}|p{1.5cm}|p{1.5cm}|p{1.5cm}|p{1.5cm}|p{1.5cm}|p{1.5cm}|}
\hline
\textbf{Model} & \textbf{Weighted F1} & \textbf{ARTHM F1} &  \textbf{LE F1} & \textbf{RENT F1} & \textbf{TimeM F1} & \textbf{TimeO F1} & \textbf{UE F1} \\
\hline
GPTLens-gte1c & 0.317 & 0.590 &  0.264 & 0.089 & 0.055 & 0.183 & 0.014 \\
\hline
GPTLens-gt1c & 0.320 & 0.601 &  0.251 & 0.092 & 0.063 & 0.187 & 0.021 \\
\hline
GPTLens-gt2c & 0.307 & 0.603 &  0.213 & 0.084 & 0.070 & 0.197 & 0.023 \\
\hline
GPTLens-gt3c & 0.317 & 0.608 &  0.201 & 0.094 & 0.045 & 0.269 & 0.025 \\
\hline
GPTLens-gt4c & 0.305 & 0.603 &  0.180 & 0.095 & 0.000 & 0.219 & 0.026 \\
\hline
GPTLens-gt5c & 0.310 & 0.609 &  0.160 & 0.095 & 0.000 & 0.250 & 0.028 \\
\hline
GPTLens-gt5f-gt5c & 0.297 & 0.608 &  0.159 & 0.095 & 0.000 & 0.095 & 0.037 \\
\hline
GPTLens-gt6c & 0.278 & 0.571 &  0.175 & 0.115 & 0.000 & 0.130 & 0.034 \\
\hline
GPTLens-gt7c & 0.200 & 0.433 &  0.153 & 0.111 & 0.000 & 0.000 & 0.000 \\
\hline
\end{tabular}
} 
\caption{F1 Scores of GPTLens analysis using GPT-4 Turbo}
\label{tab:gptlens_eval}
\end{table}

The model abbreviations shown in \cref{tab:gptlens_eval} are as follows;
\begin{itemize}
    \item \textbf{GPTLens-gte1c} - results including vulnerabilities \\
    with $correctness \geq 1$
    \item \textbf{GPTLens-gt1c} - results including vulnerabilities \\
    with $correctness > 1$
    \item \textbf{GPTLens-gt2c} - results including vulnerabilities \\
    with $correctness > 2$
    \item \textbf{GPTLens-gt3c} - results including vulnerabilities \\
    with $correctness > 3$
    \item \textbf{GPTLens-gt4c} - results including vulnerabilities \\
    with $correctness > 4$
    \item \textbf{GPTLens-gt5c} - results including vulnerabilities \\
    with $correctness > 5$
    \item \textbf{GPTLens-gt5f-gt5c} - results including vulnerabilities
     \\with $final\_score > 5$ and $correctness > 5$
    \item \textbf{GPTLens-gt6c} - results including vulnerabilities \\
    with $correctness > 6$
    \item \textbf{GPTLens-gt7c} - results including vulnerabilities \\
    with $correctness > 7$
\end{itemize}

We can see from \cref{tab:gptlens_eval} that the results are relatively similar (as measured by \textit{Weighted F1}) for a correctness score $[\geq 1, \leq 6]$.

For the rest of this paper, when we refer to \textit{GPTLens}, we are referring to the best-performing configuration from \cref{tab:gptlens_eval}, \textbf{GPTLens-gt1c}.

\subsection{Evaluation Metrics}
As there are eight potential vulnerabilities, we use a combination of metrics to evaluate how our models performed.

\subsubsection{Binary Classification}
The score is based on a binary result of whether it predicted that the smart contract had a vulnerability correctly.

\subsubsection{Classification Performance Measures}
We use the calculated Accuracy, Precision, Recall, and F1 Score to evaluate the models' performance.

We also take a weighted F1 Score to measure the effectiveness overall.

\section{Results analysis}\label{sec:analysis}
In the following tables the models and vulnerabilities are largely represented as abbreviations.

\subsection{Abbreviation guide}
The names included in the tables are listed below.
\subsubsection{Models}
\begin{itemize}
    \item \textbf{DL-Foundation} - \textit{Detect Llama - Foundation} - this model was fine-tuned on the full 17,000 record dataset and uses Meta's 34b parameter Code Llama Foundation model\cite{roziere_code_2023}
    \item \textbf{DL-Instruct} - \textit{Detect Llama - Instruct} - this model was also fine-tuned on the full dataset; however, it uses the Instruct trained variant of Meta's 34b parameter Code Llama model\cite{roziere_code_2023}
    \item \textbf{GPT-4} - \textit{GPT-4 Zero-shot Analysis} - OpenAI's GPT-4 Model\cite{openai_gpt-4_2023} with a specific prompt identifying what to look for (seen in listing \ref{lst:gpt4_prompt}) using the function calling feature\cite{eleti_function_2023} to structure the data.
    \item \textbf{GPT-4 Turbo} - \textit{GPT-4 Turbo Zero-shot Analysis} - OpenAI's GPT-4 Turbo Model\cite{openai_new_2023} with a specific prompt identifying what to look for (seen in listing \ref{lst:gpt4_prompt}) using the function calling feature\cite{eleti_function_2023} to structure the data.
    \item \textbf{GPT-4 Critic} - \textit{GPT-4 with Critic Step from \cite{hu_large_2023}} - results from GPT-4 processed using an additional critic analysis step using listing \ref{lst:gpt4_critic} and \ref{lst:gpt4_critic_vuln}.
    \item \textbf{GPT-4 Turbo Critic} - \textit{GPT-4 Turbo with Critic Step from \cite{hu_large_2023}} - results from GPT-4 Turbo processed using an additional critic analysis step using listing \ref{lst:gpt4_critic} and \ref{lst:gpt4_critic_vuln}.
    \item \textbf{GPT-3.5FT} - \textit{GPT-3.5 Turbo Fine-tune} - OpenAI's GPT-3.5 Turbo\cite{peng_gpt-35_2023} fine-tuned with the 4,000 record detection dataset.
    \item \textbf{GPTLens} - \textit{Best performing GPTLens\cite{hu_large_2023} ranking} - the best performing ranking from \cref{tab:gptlens_eval}.
    \item \textbf{Random} - \textit{Random baseline} - a randomly generated baseline for comparison.
\end{itemize}

\subsubsection{Vulnerabilities} originally from \cite{yashavant_scrawld_2022} by Yashavant et al.
\begin{itemize}
    \item \textbf{LE} - \textit{Locked Ether}
    \item \textbf{ARTHM} - \textit{Arithmetic (Integer Overflow and Underflow)}
    \item \textbf{DOS} - \textit{Denial of Service}
    \item \textbf{RENT} - \textit{Reentrancy}
    \item \textbf{TimeM} - \textit{Time Manipulation (Block values as a proxy for time)}
    \item \textbf{TimeO} - \textit{Timestamp Ordering (Transaction Order Dependence)}
    \item \textbf{Tx-Origin} - \textit{Authorization through tx.origin}
    \item \textbf{UE} - \textit{Unhandled Exception (Unchecked Call Return Value)}
\end{itemize}

\begin{table}
\centering
  \caption{Binary Vulnerability Classification results}
  \label{tab:binary}
  \begin{tabular}{|p{1.75cm}|c|c|c|c|c|c|}
    \hline
    \textbf{Model} & \textbf{Precision} & \textbf{Recall}& \textbf{F1} & \textbf{Specificity} & \textbf{Accuracy}\\
    \hline
    DL-Foundation & 0.517 & 0.993 & 0.68 & 0.023 & 0.521 \\
    \hline

    DL-Instruct & 0.774 & 0.443 & 0.563 & 0.864 & 0.648 \\
    \hline
    GPT-4 & 0.675 & 0.646 & 0.66 & 0.676 & 0.661  \\
    \hline

    GPT-4 Critic & 0.679 & 0.635 & 0.656 & 0.688 & 0.661 \\
    \hline

    GPT-4 Turbo & 0.629 & 0.727 & 0.675 & 0.549 & 0.640 \\
    \hline
    GPT-4 Turbo Critic & 0.623 & 0.646 & 0.634 & 0.588 & 0.617 \\
    \hline
    GPT-3.5FT & 0.77 & 0.782 & 0.776 & 0.77 & 0.776 \\
    \hline  
    GPTLens* & 0.533 & 0.988 & 0.692 & 0.147 & 0.564 \\
    \hline 
    Random & 0.508 & 0.79 & 0.618 & 0.195 & 0.5 \\
    \hline
\end{tabular}
\end{table}

\subsection{RQ1: How effective is GPT-4 at zero-shot vulnerability detection?}
We can see from our results in \cref{tab:binary} that for binary classification, GPT-4 (and GPT-4 Turbo) achieves an F1 score of slightly better than random, moderately better than DL-Instruct, similar to DL-Foundation and moderately worse than GPT-3.5FT.

However, as random performs relatively well in \cref{tab:binary}, it is not the best measure for us to use.

If we look at \cref{tab:overall_scores}, we can use the weighted F1 - a score from sklearn.metrics that uses the number of True Positive values for each label classification to weight the F1 score\cite{scikit-learn} - as a general guide to the effectiveness of a model.

We can see that, generally, GPT-4 and GPT-4 Turbo perform only slightly better than random in identifying the eight vulnerabilities, slightly worse than DL-Instruct overall and significantly worse than DL-Foundation and GPT-3.5FT models. However, GPT-4 performs only slightly worse than the best performer, GPT-3.5FT, in identifying the Arithmetic vulnerability in smart contracts (as shown in \cref{tab:overall_scores}).

\begin{table}
  \caption{F1 Scores for all models and all vulnerabilities}
  \label{tab:overall_scores}
  \resizebox{\textwidth}{!}{%
\begin{tabular}{|p{2.2cm}|p{1.5cm}|p{1.5cm}|p{1.2cm}|p{1.2cm}|p{1.2cm}|p{1.2cm}|p{1.2cm}|p{1.2cm}|p{1.2cm}|p{1.2cm}|p{1.2cm}|}
\hline
    \textbf{Model} & \textbf{Weighted F1} & \textbf{ARTHM F1} & \textbf{DOS F1} & \textbf{LE F1} & \textbf{RENT F1} & \textbf{TimeM F1} & \textbf{TimeO F1} &	\textbf{Tx-Origin F1} & \textbf{UE F1}\\
    \hline
    GPT-3.5FT & 0.61 & 0.639 & 0 & 0.81 & 0.185 & 0 & 0.219 & 0 & 0 \\
    \hline
    random & 0.184 & 0.268 & 0 & 0.188 & 0.106 & 0.042 & 0.222 & 0 & 0 \\
    \hline
    DL-Foundation & 0.568 & 0.493 & 0 & 0.36 & 0.048 & 0 & 0.174 & 0 & 0 \\
    \hline
    DL-Instruct & 0.297 & 0.517 & 0 & 0.269 & 0.056 & 0 & 0.175 & 0 & 0 \\
    \hline
    GPT-4 & 0.218 & 0.609 & 0 & 0 & 0.1 & 0 & 0.17 & 0 & 0.02 \\
    \hline
    GPT-4 Turbo & 0.243 & 0.593 & 0 & 0.133 & 0.073 & 0.070 & 0.172 & 0 & 0 \\
    \hline
    GPT-4 Critic & 0.226 & 0.586 & 0 & 0.101 & 0 & 0.137 & 0 & 0 & 0 \\
    \hline
    GPT-4 Turbo Critic & 0.255 & 0.591 & 0.075 & 0.086 & 0.123 & 0.193 & 0 & 0 & 0 \\
    \hline
    GPTLens* & 0.320 & 0.601 & 0.251 & 0.092 & 0.063 & 0.187 & 0 & 0.021 & 0 \\
    \hline
  \end{tabular}
  } 
\end{table}

\subsection{RQ2: Can we fine-tune an open-source model to be more effective than GPT-4?}
As discussed earlier in our paper, we fine-tuned two variants of Meta's Code Llama model, Detect Llama (DL) - Foundation and DL Instruct.

For binary classification (as seen in \cref{tab:binary}), we can see that the DL-Foundation model performs similarly to GPT-4 and GPT-4 Turbo and slightly better than random, whereas the DL-Instruct model scores moderately worse than random and the GPT-4 models when comparing F1 scores.

However, when we examine the weighted F1 scores in \cref{tab:overall_scores}, we can see that DL-Instruct moderately outperforms the GPT-4 models and random, whereas DL-Foundation significantly outperforms random, the GPT-4 models and DL-Instruct with a weighted F1 of $0.568$.

\subsection{RQ3: Can we fine-tune GPT-3.5 Turbo to be more effective than GPT-4?}
We can see from both \cref{tab:binary} and \cref{tab:overall_scores} that our fine-tuned GPT-3.5 Turbo is at least moderately better than all of the other models at binary classification, and for general performance (using weighted F1 as a guide) performs slightly better on average than our DL-Foundation model, and significantly better than our DL-Instruct model, the GPT-4 models and random.

\subsection{RQ4: How effective is our model when compared to alternate vulnerability detection techniques using GPT-4?}
To evaluate against other techniques, we focus on a modified GPTLens\cite{hu_large_2023} using GPT-4 Turbo Preview for Auditor and Critic, as well as an additional critic step applied to the GPT-4 and GPT-4 Turbo results.

We can see from \cref{tab:overall_scores} that our modified GPTLens outperforms (based on weighted F1 score) both GPT-4 and GPT-4 Turbo and our DL-Instruct model. However, GPTLens significantly under-performs our DL-Foundation model and the GPT-3.5FT model with a weighted F1 of $0.320$ for GPTLens*, $0.568$ for DL-Foundation and $0.61$ for GPT-3.5FT.

GPT-4 Critic and GPT-4 Turbo Critic see only a slight increase in performance over the models without the critic step (weighted F1 score of $0.218$ vs $0.226$ for GPT-4 and GPT-4 Critic and $0.243$ and $0.255$ for GPT-4 Turbo and GPT-4 Turbo Critic respectively).

\subsection{Further analysis}
If we further examine the results in \cref{tab:overall_scores}, we can see that the only vulnerabilities where many models outperform random by a significant amount are ARTHM, or Arithmetic, and LE, or Locked Ether.

To further identify the accuracy of the models over those two vulnerabilities, we can view the results in further detail in \cref{tab:ar_le_scores}.

\begin{table}
  \caption{Scores for ARTHM and LE Vulnerabilities}
  \label{tab:ar_le_scores}
  \resizebox{\textwidth}{!}{%
  \begin{tabular}{|p{2.2cm}|p{1.5cm}|p{1.5cm}|p{1.2cm}|p{1.2cm}|p{1.2cm}|p{1.2cm}|p{1.2cm}|p{1.2cm}|p{1.2cm}|p{1.2cm}|p{1.2cm}|}

  \hline
    \textbf{Model} & \textbf{Weighted F1} & \textbf{ARTHM Prec.} & \textbf{ARTHM Recall} & \textbf{ARTHM F1} & \textbf{ARTHM Acc.} & \textbf{LE Prec.} & \textbf{LE Recall} & \textbf{LE F1} & \textbf{LE Acc.}\\
    \hline
    GPT-3.5FT & 0.719 & 0.65 & 0.63 & 0.639 & 0.77 & 0.823 & 0.798 & 0.81 & 0.926 \\
    \hline
    random & 0.225 & 0.311 & 0.235 & 0.268 & 0.584 & 0.168 & 0.212 & 0.188 & 0.636 \\
    \hline
    DL-Foundation & 0.674 & 0.336 & 0.92 & 0.493 & 0.386 & 0.625 & 0.253 & 0.36 & 0.822 \\
    \hline
    DL-Instruct & 0.350 & 0.586 & 0.463 & 0.517 & 0.72 & 0.8 & 0.162 & 0.269 & 0.826 \\
    \hline
    GPT-4 & 0.363 & 0.652 & 0.571 & 0.609 & 0.763 & 0 & 0 & 0 & 0.785 \\
    \hline
    GPT-4 Turbo & 0.429 & 0.550 & 0.642 & 0.593 & 0.714 & 0.148 & 0.121 & 0.133 & 0.688 \\
    \hline
    GPT-4 Critic & 0.338 & 0.659 & 0.528 & 0.586 & 0.759 & 0 & 0 & 0 & 0.795 \\
    \hline
    GPT-4 Turbo Critic & 0.393 & 0.584 & 0.599 & 0.591 & 0.732 & 0.143 & 0.051 & 0.075 & 0.752 \\
    \hline
    GPTLens* & 0.441 & 0.645 & 0.562 & 0.601 & 0.758 & 0.261 & 0.242 & 0.251 & 0.714 \\
    \hline
  \end{tabular}
  } 
\end{table}

In \cref{tab:ar_le_scores}, we can see that the GPT-4's performance has increased to be significantly above random and slightly above DL-Instruct (and GPT-4 Turbo performing moderately better than DL-Instruct), and DL-Foundation and GPT-3.5FT have increased their weighted F1 score to $0.674$ and $0.719$ respectively. We can also see in \cref{tab:ar_le_scores} that GPTLens* performs slightly better than GPT-4 Turbo, however, the GPT-4 models with an additional critic step perform slightly worse than the GPT-4 models individually.

The downward performance trend of the GPT-4 models with critic in \cref{tab:ar_le_scores} is likely due to the increase in performance in by GPT-4 Critic and GPT-4 Turbo Critic models at identifying TimeM vulnerability than the original models (as shown by the TimeM F1 Score in \cref{tab:overall_scores}).

\section{Discussions}\label{sec:discuss}
In this section, we discuss improvements that can be made to our models and future work.

\subsection{Increasing Solidity version range}
As we mentioned earlier in our paper, due to the age of the tools used in \cite{yashavant_scrawld_2022}, all of the smart contracts in our test set had to be Solidity version $0.4.x$. The current version of Solidity is $0.8.22$\cite{solidity_team_solidity_2023}, so for the tool to be as accurate and useful in wide, general release we could update the tools used for the majority vote to support later versions of Solidity.

This would allow us to create a new training set with smart contracts from Solidity version $0.8.x$.

\subsection{Focusing vulnerability detection}
As we are searching for eight different vulnerabilities with varying levels of success and accuracy (as seen in \cref{tab:overall_scores}), we could improve results with less well-detected vulnerabilities by identifying more smart contracts that had only those vulnerabilities and adding them to the training set.

\subsection{Reducing model size}
The Llama Code base models from Meta that were used for fine-tuning of our models are 34 billion parameters. The 34b parameter models are the largest; Meta also released 13 billion and 7 billion parameter models of the Foundation and Instruct variants used for training\cite{roziere_code_2023}. To serve our 34b parameter Detect Llama models with the popular Text Generation Inference engine from Huggingface\cite{huggingface_text_2023} requires a single A100 80gb GPU.

For future research, we could train smaller models with a lower parameter count to see how much accuracy is lost. If a smaller model can provide a similar amount of accuracy once trained, it would make it faster, cheaper and more accessible to run.

\section{Conclusion}\label{sec:concl}
In this work, we introduce our two trained open-source models, Detect Llama - Foundation and Detect Llama - Instruct; fine-tuned versions of Meta's Code Llama\cite{roziere_code_2023} 34b Foundation and Instruct models, respectively.

We then evaluate these models against a fine-tuned version of GPT-3.5 Turbo and OpenAI's GPT-4 and GPT-4 Turbo Preview.

We find that on a weighted F1 score of all eight vulnerabilities and two best-predicted vulnerabilities (across all models), our Detect Llama - Foundation model significantly outperformed GPT-4 and GPT-4 Turbo, with our model scoring weighted F1 of $0.568$ and $0.674$ respectively compared to GPT-4's $0.218$ and $0.363$, and GPT-4 Turbo's $0.243$ and $0.429$.

One surprise we found from our research was that our fine-tuned GPT-3.5 Turbo model outperformed all other models. Achieving a weighted F1 score of $0.61$ for all vulnerabilities and $0.719$ for the two best-detected vulnerabilities. The performance of the fine-tuned GPT-3.5 Turbo model was surprising, as the fine-tuning process is not listed as adding new data or abilities but rather  \textit{Improved steerability, reliable output formatting and custom tone}\cite{peng_gpt-35_2023}.

This research also releases our two open-source models, Detect Llama - Foundation\cite{ince_detect_2023} and Detect Llama - Instruct\cite{ince_detect_2023-1}, and the training\cite{ince_smart_2023} and evaluation\cite{ince_peterdouglasdetect-llama-evaluation_2024} datasets; aiding to lay the groundwork for further research into the area of Large Language Models for smart contract vulnerability detection.

\subsubsection*{Acknowledgements.} This paper is supported by Australian Research Council (ARC) Discover Project DP220101234, partially supported by ARC under project DE210100019 and Collaborative research project (H-ZGGQ). 

%
%
%
\bibliographystyle{splncs04}
\bibliography{detect-llama}


\end{document}